\def\vec#1{{\rm\bf #1}}

\documentstyle[aps,multicol,epsfig]{revtex}

\begin{document}

\title{Statistical Mechanics of Stress Transmission in Disordered Granular Arrays}
\author{S. F. Edwards and  D. V. Grinev}
\address{Cavendish Laboratory, University of Cambridge, Madingley Road, Cambridge CB3 OHE, United Kingdom}
\date{\today}
\maketitle

\begin{abstract}
We give a statistical-mechanical theory of stress transmission in disordered arrays of rigid grains with perfect friction. Starting from the equations of microscopic force and torque balance we derive the fundamental equations of stress equilibrium. We illustrate the validity of our approach by solving the stress distribution of a homogeneous and isotropic array.
\end{abstract}

\multicols{2}

Transmission of stress and statistics of force fluctuations in static granular arrays are fundamental, but unresolved problems in physics \cite{Jaeger,stress}. Despite several theoretical attempts \cite{Wittmer,Coppersmith} and a vast engineering literature \cite {Nedderman,Wood} the connectivity of granular media is still poorly understood at a fundamental level.
In this Letter we propose  a theory of stress transmission in disordered arrays of rigid cohesionless grains with perfect friction. A real granular aggregate (e.g. sand or soil) is a very complex object \cite{Wood}. However, simple models are easier to comprehend, and extra complexities can always be incorporated subsequently. In our case the rigid grain paradigm provides a crucial starting point from which to appreciate the theoretical physics of the problem. 
We model the granular material as an assembly of discrete rigid particles whose interactions with their neighbours are localized at pointlike contacts. Therefore the description of the network of intergranular contacts is essential for the understanding of force transmission in granular assemblies. 
Grain $\alpha$ exerts a force on grain $\beta$ at a point $ {\vec {\cal R}}^{\,\alpha\beta}\,=\,\vec{R}^{\,\alpha}+\vec{r}^{\,\alpha\beta}$. The contact is a point in a plane whose normal is $\vec{n}^{\,\alpha\beta}$. The vector $\vec{R}^{\,\alpha}$is defined by:

\begin{equation}\label{frank1}
\vec{R}^{\,\alpha}=\frac{\sum_{\,\beta} \,\vec{{\cal R}}^{\,\alpha\beta}}{z}\, ,
\end{equation}
so that $\vec{R}^{\,\alpha}$ is the centroid of contacts, and hence

\begin{equation}\label{frank2}
\sum_{\beta} \vec{r}^{\,\alpha\beta}\,=\,0\, ,\quad \vec{R}^{\,\alpha\beta}=\vec{r}^{\,\alpha\beta}- \vec{r}^{\,\beta\alpha}\, ,
\end{equation}
where $z$ is the number of contacts per grain and $\sum_{\,\beta}$ means summation over the nearest neighbours. Hence $\vec{R}^{\,\alpha}$, $\vec{r}^{\,\alpha\beta}$ and $\vec{n}^{\,\alpha\beta}$ are geometrical properties of the aggregate under consideration and the other shape specifications do not enter.
Friction is assumed to be infinite and the geometry is frozen after the deposition and can not be changed by applying or removing an external force on the boundaries.
In a static array Newton's equations of integranular force and torque balance are satisfied.
Balance of force around the grain $\alpha$ requires

\begin{equation}\label{frank3}
\sum_{\beta}f^{\,\alpha\beta}_{i}=g^{\alpha}_{i}\, ,
\end{equation}
\begin{equation}\label{frank4}
f^{\,\alpha\beta}_{i}+f^{\,\beta\alpha}_{i}=0\, ,
\end{equation}
where $i=1,2,3$ are cartesian indices and ${\vec g}^{\,\alpha}$ is the external force acting on grain $\alpha$. Further on in this Letter ${\vec g}$ is used also for the external forces at the boundaries.

The equation of torque balance is

\begin{equation}\label{frank5}
\sum_{\beta}\epsilon_{ikl}\,f^{\,\alpha\beta}_{k}\,r^{\,\alpha\beta}_{l}=C_{i}^{\,\alpha}\, .
\end{equation}

The centroid of the contact points need not coincide with the centroid of the forces e.g. the centre of mass of a solid grain, but we will assume it is so in order to keep the analysis simple so that we ensure that the macroscopic stress tensor is symmetric, at least on average. 
It can be verified that, for the integranular forces in the static array to be determined by these equations,  the coordination number $z=3$ in 2-D and $z=4$ in 3-D is required. In this paper we present the results for the 2-D case only. 
The microscopic version of stress analysis is to determine all of the intergranular forces, given the applied force,
torque loadings on each grain and geometric specification of a granular array.
The number of unknowns per grain is $zd/2$. Required force and torque equations give $d + \frac{d(d-1)}{2}$ constraints.
The system of equations for the integranular forces is complete when the coordination number is  $z_m \,=\,d+1$.
Theory which confirms this observation has been proposed for periodic arrays of grains with perfect and zero friction \cite{Grinev}. It is clear that the coordination number  $z$ controls the connectivity of granular media. We will assume that $z$ is indeed 3 in 2-D, for this is surely the simplest situation, and one which is physically possible.
 The ultimate goal, however, is to determine the macroscopic stress tensor at every point of a granular array, given external loadings and geometric specification.
The macroscopic state of stress is a function of the distribution of contact forces. For any aggregate of discrete grains subjected to external loading, the transmission of stress from one point to another can only occur via the intergranular contacts. Therefore it is clear that the network of contacts determines the distribution of stresses within the granular array. The network of contacts is determined by the deposition history of the sample and the external loading on the boundaries.
We define the tensorial force moment:

\begin{equation}\label{frank6}
S^{\,\alpha}_{ij}=\sum_{\beta}f^{\,\alpha\beta}_{i}\,r^{\,\alpha\beta}_{j}\, ,
\end{equation}
which is the microscopic analogue of the stress tensor. With $C_{i}^{\,\alpha}=0$, $S^{\,\alpha}_{ij}$ will be symmetric.
 Our goal is to find a complete system of equations for the macroscopic stress tensor $\sigma_{ij}$, which is supported by the given network of contacts in the state of mechanical equilibrium. 
Given an assembly of discrete grains which is represented by a very complex network of contacts, we associate a continuous medium to have continuously distributed properties. Such spatial smoothing or coarse-graining can be accomplished formally.
To obtain the macroscopic stress tensor from the tensorial force moment to the macroscopic stress tensor we coarse-grain i.e. average it over an ensemble of configurations:

\begin{equation}\label{frank7}
\sigma_{ij}({\vec r})\,=\,\langle \,\sum_{\alpha=1}^{N}\,S_{ij}^{\,\alpha}\, \delta({\vec r}-{\vec R^{\,\alpha}})\,\rangle\, .
\end{equation}

In the simplest cases of isotropic and homogeneous arrays this is not a problem. The difficulties appear when the array under consideration is anisotropic or inhomogeneous. Within the confines of this paper we explore only the simplest cases.
The number of equations required equals the number of independent components of a symmetric stress tensor $\sigma_{ij}\,=\,\sigma_{ji}$ and is $\frac{d\,(d\,+\,1)}{2}$. At the same time, the number of equations available is $d$. These are vector equations of the stress equilibrium  $\frac{\partial \sigma_{ij}}{\partial x_{j}}\,=\,g_{i}$ which have their origin in Newton's second law. Therefore we have to find $\frac{d(d-1)}{2}$ equations, which possess the information from Newton's third law, to complete and solve the system of equations which governs the transmission of stress in a granular array. Thus in 2-D there is one missing equation, and we derive it in terms of the geometry of the system. 

Given the set of equations $(\ref{frank3}-\ref{frank5})$ we can write the probability functional for the integranular force  $f_{i}^{\,\alpha\beta}$ as

\begin{eqnarray}\label{frank8}
 P \left \{ f_{i}^{\,\alpha\beta} \right \} &=&{\cal N}
\delta\big(\sum_{\beta}f^{\,\alpha\beta}_{i}-g^{\alpha}_{i}\big) \nonumber \\
& &\mbox{}\times
\delta\big(\sum_{\beta}\epsilon_{ikl}\,f^{\,\alpha\beta}_{k}\,r^{\,\alpha\beta}_{l}\big) \nonumber \\
& &\mbox{}\times
\delta\big(f^{\,\alpha\beta}_{i}+f^{\,\beta\alpha}_{i}\big)\,,
\end{eqnarray}
where the normalization, ${\cal N}$, which is a function of a configuration, is

\begin{eqnarray}\label{frank9}
{\cal N}^{-1}&=&\int\prod_{\alpha,\beta}\, P \left \{f_{i}^{\,\alpha\beta} \right\}\,{\cal D}f^{\,\alpha\beta}\, .
\end{eqnarray}

The probability of finding the tensorial force moment $S_{ij}^{\,\alpha}$ on grain $\alpha$ is

\begin{eqnarray}\label{frank10}
P \left \{ S_{ij}^{\,\alpha} \right \} &=& \int\prod_{\alpha,\beta}
\delta\Big (S^{\,\alpha}_{ij}-\sum_{\beta}f^{\,\alpha\beta}_{i}r^{\,\alpha\beta}_{j}\Big ) \,P \left \{ f_{i}^{\,\alpha\beta} \right \} \,{\cal D}f^{\,\alpha\beta}
\end{eqnarray}
where $\int\,{\cal D}f^{\,\alpha\beta}$ implies integration over all functions $f^{\,\alpha\beta}$, since all the constraints on $f^{\,\alpha\beta}$ have been experienced. We assume that the $z=d+1$ condition means that the integral exists.

We exponentiate the delta functions and thus introduce the set of conjugate fields $\zeta^{\,\alpha}_{ij}$, $\gamma^{\,\alpha}_{i}$, $\lambda^{\,\alpha}_{i}$ and $\eta^{\,\alpha\beta}_{i}$.

\begin{equation}\label{frank11}
P \left \{ S_{ij}^{\,\alpha} \right \}=\int\prod\,e^{iA}\,{\cal D}f^{\,\alpha\beta}\,{\cal D}\zeta^{\,\alpha}\,{\cal D}\gamma^{\,\alpha}\,{\cal D}\lambda^{\,\alpha}\,{\cal D}\eta^{\,\alpha\beta}\, ,
\end{equation}
where $A$ is

\begin{eqnarray}\label{frank12}
\lefteqn{A\,=\,\sum_{\alpha}\,\zeta^{\,\alpha}_{ij}\Big (S^{\,\alpha}_{ij}-\sum_{\beta}f^{\,\alpha\beta}_{i}r^{\,\alpha\beta}_{j}\Big )} \nonumber \\
& &\mbox{}+
\gamma^{\,\alpha}_{i}\Big(\sum_{\beta}f^{\,\alpha\beta}_{i}-g^{\,\alpha}_{i}\Big)\nonumber  \\
& &\mbox{}+
\lambda^{\,\alpha}_{i}\Big(\sum_{\beta}\epsilon_{ikl}f^{\,\alpha\beta}_{k}r^{\,\alpha\beta}_{l}\Big ) \nonumber \\
& &\mbox{}+
\eta^{\,\alpha\beta}_{i}\Big(f^{\,\alpha\beta}_{i}\,+\,f^{\,\beta\alpha}_{i}\Big)\, . 
\end{eqnarray}
The $\lambda^{\,\alpha}$ field term gives the symmetry of $S^{\,\alpha}_{ij}$.
After integrating out the $f^{\,\alpha\beta}$ and $\eta^{\,\alpha\beta}$ fields we find the following linear equation for the conjugate fields:

\begin{equation}\label{frank13}
\zeta^{\,\alpha}_{ij}r^{\,\alpha\beta}_{j}-\gamma^{\,\alpha}_{i}=\zeta^{\,\beta}_{ik}r^{\,\beta\alpha}_{k}-\gamma^{\,\beta}_{i}.
\end{equation}

The idea of the conjugate fields method is to use these equations for the $\zeta$ field in the stress probability functional, in order to derive the complete system of equations for the stress tensor.
The general solution of the above equation is a sum of the $\zeta^{0}$ field which is the particular solution and depends on $\gamma$,  and $\zeta^{*}$ which is the complimentary function

\begin{equation}\label{frank14}
\zeta^{\,\alpha}_{ij}\,=\,\zeta^{\,\alpha\,0}_{ij}\,+\,\zeta^{\,\alpha\,*}_{ij}\, .
\end{equation}

If we introduce the fabric tensor $F^{\,\alpha}_{ij}$ and its inverse $ M^{\,\alpha}_{ij}$:

\begin{equation}\label{frank15}
F_{ij}^{\,\alpha}\,=\,\sum_{\beta}\,R^{\,\alpha\beta}_{i}\,R^{\,\alpha\beta}_{j}\, ,\quad
M_{ij}^{\,\alpha}\,=\,\Big(F^{\,\alpha}\Big)^{-1}_{ij}\, ,
\end{equation}
we can rewrite the equation (\ref{frank13}) in the following form:

\begin{equation}\label{frank16}
\zeta^{\,\alpha}_{ij}=M^{\,\alpha}_{jl}\sum_{\beta}R^{\,\alpha\beta}_{l}(\gamma^{\,\alpha}_{i}-\gamma^{\,\beta}_{i})+M^{\,\alpha}_{jl}\sum_{\beta}R^{\,\alpha\beta}_{l}r^{\,\beta\alpha}_{k}(\zeta^{\,\beta}_{ik}-\zeta^{\,\alpha}_{ik})\, .
\end{equation}
which permits an expansion based on the first two terms, i.e:
\begin{eqnarray}\label{frank16a}
\lefteqn{\zeta^{\,\alpha}_{ij} \simeq M^{\,\alpha}_{jl}\sum_{\beta}R^{\,\alpha\beta}_{l}(\gamma^{\,\alpha}_{i}-\gamma^{\,\beta}_{i})\,+} \nonumber \\
& &\mbox{}+
M^{\,\alpha}_{jl}\sum_{\beta}R^{\,\alpha\beta}_{l}r^{\,\beta\alpha}_{k}M^{\,\beta}_{km}\sum_{\delta}R^{\,\beta\delta}_{m}\,(\gamma^{\,\beta}_{i}\,-\,\gamma^{\,\delta}_{i})+...
\end{eqnarray}

\underline{\sc The stress-force equation}
The next step is to integrate out the $\gamma$ field which gives us the stress-force equation:
                           
\begin{equation}\label{frank17}
\sum_{\beta}M^{\,\alpha}_{jl} R^{\,\alpha\beta}_{l}S^{\,\alpha}_{ij}\,-\,\sum_{\beta} M^{\,\beta}_{jl} R^{\,\beta\alpha}_{l}S^{\,\beta}_{ij}\,=\,g^{\,\alpha}_{i}\, .
\end{equation}
So by expanding $\beta$ quantities about $\alpha$ quantities we reach:

\begin{equation}\label{frank18}
\nabla_{j}\, \sigma_{ij}\,+\,\nabla_{j}\nabla_{k}\nabla_{m}\,K_{ijkl}\sigma_{lm}\,+ ...=\,g_{i}.
\end{equation}
where $K_{ijkl} = \langle \, R_{i}^{\,\alpha\beta}\,R_{j}^{\,\alpha\beta}\,R_{k}^{\,\alpha\beta}\,R_{l}^{\,\alpha\beta}\,\rangle$ and gives a correction to the standard equation of stress equilibrium  $\nabla_{j}\, \sigma_{ij}=g_{i}$ at the length-scale which is small compared to the size of the system. These corrections correspond to the presence of the second, third etc. nearest neighbours and topological correlations and must vanish in the  $k\,\rightarrow \,0$ limit.

\underline{\sc The stress-geometry equation} So far the well-known equations have been derived by using the information from Newton's second law. But we still have unused information from Newton's third law. By integrating out the $\zeta^{\,*}$ field we obtain the missing equations we are looking for.
Let us consider that part of the eqn.(\ref{frank11}) which contains the $\zeta^{\,*}$ field:

\begin{equation}\label{frank19}
\int \,e^{i\,\sum_{\alpha=1}^{N}\,\zeta^{\,\alpha\,*}_{ij}\,S_{ij}^{\,\alpha}}\,\delta(\zeta^{\,\alpha\,*}_{ij}r^{\,\alpha\beta}_{j}\,-\,\zeta^{\,\beta\,*}_{ij}r^{\,\beta\alpha}_{j})\,\prod_{\alpha=1}^{N}\,{\cal D}\zeta^{\,\alpha\,*}_{ij}\, ,
\end{equation}
and
\begin{equation}\label{frank20}
\zeta^{\,\alpha\,*}_{ij}r^{\,\alpha\beta}_{j}\,-\,\zeta^{\,\beta\,*}_{ij}r^{\,\beta\alpha}_{j} = 0\, .
\end{equation}
Counting the degrees of freedom in this equation we note that it can only give two (scalar) equations in 2-D and three in 3-D.
Using ${\vec R}^{\,\alpha\beta}$ and ${\vec Q}^{\,\alpha\beta}\,=\,{\vec r}^{\,\alpha\beta}\,+\,{\vec r}^{\,\beta\alpha}$ , we can get these equations by projecting the vector equation equation into:

\begin{equation}\label{frank21}
\zeta^{\,\alpha\,*}_{ij}\sum_{\beta}\,R^{\,\alpha\beta}_{i}R^{\,\alpha\beta}_{j}\,+\,\sum_{\beta}(\zeta^{\,\alpha\,*}_{ij}-\,\zeta^{\,\beta\,*}_{ij})r^{\,\beta\alpha}_{j}R^{\,\alpha\beta}_{i}=0 \, , 
\end{equation}
\begin{equation}\label{frank22}
\zeta^{\,\alpha\,*}_{ij}\sum_{\beta}\,Q^{\,\alpha\beta}_{i}R^{\,\alpha\beta}_{j}\,+\,\sum_{\beta}(\zeta^{\,\alpha\,*}_{ij}-\,\zeta^{\,\beta\,*}_{ij})r^{\,\beta\alpha}_{j}Q^{\,\alpha\beta}_{i}=0\, ,
\end{equation}
It should be emphasised that the system under consideration is disordered and therefore ${\vec Q}^{\,\alpha\beta}\,\neq \,0$ (whereas for a honeycomb periodic array ${\vec Q}^{\,\alpha\beta}\,= \,0$). Assuming as before that $\zeta^{\,\alpha\,*}_{ij}-\,\zeta^{\,\beta\,*}_{ij}$ gives rise to gradient terms we can exponentiate (\ref{frank21}) and (\ref{frank22}) by parametric variables $\phi^{\,\alpha}$ and $\psi^{\,\alpha}$:

\begin{equation}\label{frank23}
\int \,e^{i\,\sum_{\alpha=1}^{N}\,\zeta^{\,\alpha\,*}_{ij}\,(S_{ij}^{\,\alpha}-\phi^{\,\alpha}F^{\,\alpha}_{ij}-\psi^{\,\alpha}G^{\,\alpha}_{ij})}\,\prod_{\alpha}^{N}\,{\cal D}\zeta^{\,\alpha\,*}_{ij}\,{\cal D}\phi^{\,\alpha}{\cal D}\psi^{\,\alpha}\, .
\end{equation}
where $F_{ij}^{\,\alpha}$ is given by (\ref{frank15}) and 
\begin{equation}\label{frank24}
G_{ij}^{\,\alpha}\,=\,\frac{1}{2}\Big(\sum_{\beta}\,Q^{\,\alpha\beta}_{i}\,R^{\,\alpha\beta}_{j}+Q^{\,\alpha\beta}_{j}\,R^{\,\alpha\beta}_{i}\Big)\, .
\end{equation}
After integrating out the $\zeta^{\,\alpha\,*}$, $\phi^{\,\alpha}$ and $\psi^{\,\alpha}$ fields, we find the following equation for $S_{ij}^{\,\alpha}$:

\begin{equation}\label{frank25}
\left| \begin{array}{ccc}
              S^{\,\alpha}_{11} &  F_{11}^{\,\alpha} & G_{11}^{\,\alpha}    \\
                                                                                                           
              S^{\,\alpha}_{22} &  F_{22}^{\,\alpha} & G_{22}^{\,\alpha}   \\

              S^{\,\alpha}_{12} &  F_{12}^{\,\alpha} & G_{12}^{\,\alpha}
\end{array} \right|=0\, .
\end{equation}

Note that although there are explicit forms generalising (\ref{frank25}) in 3-D, these are more complex algebraically as a consequence of the higher coordination number. $ F_{ij}^{\,\alpha}$ and $G_{ij}^{\,\alpha}$ will depend on configuration and averaging (\ref{frank25}) is quite complex. 
The simplest array will have ${\vec Q}^{\,\alpha\beta}$ orthogonal to ${\vec R}^{\,\alpha}$, i.e. if ${\vec R}^{\,\alpha\beta}\,~=~(X^{\,\alpha\beta},Y^{\,\alpha\beta})$, then ${\vec Q}^{\,\alpha\beta}\,~=~(Y^{\,\alpha\beta},-X^{\,\alpha\beta})$. It follows, that $F_{ij}^{\,\alpha}$ and $ G_{ij}^{\,\alpha}$ can be written as
\begin{equation}\label{frank26}
 \ F_{ij}^{\,\alpha}\, = \,\left( \begin{array}{cc}
              1 & 0 \\
                                                                                                           \\
              0 & 1 
\end{array} \right)\, ,\qquad
\ G_{ij}^{\,\alpha}\, = \,\left( \begin{array}{cc}
              \mbox{sin} \theta^{\,\alpha} & \mbox{cos} \theta^{\,\alpha} \\
                                                                                                           \\
              \mbox{cos} \theta^{\,\alpha} & -\mbox{sin} \theta^{\,\alpha}
\end{array} \right)\, .
\end{equation}
Then eqn.(\ref{frank25}) can be rewritten:
\begin{equation}\label{frank27}
S^{\,\alpha}_{22}\,-\,S^{\,\alpha}_{11}\,=\,2S^{\,\alpha}_{12}\,\mbox{tan}\,\theta^{\,\alpha}
\end{equation}
Thus if we are given $S_{12}^{\,\alpha}$, the probability of finding $S_{11}^{\,\alpha}\,-\,S_{22}^{\,\alpha}$ is
\begin{equation}\label{frank28}
P \left \{S_{11}^{\,\alpha}\,-\,S_{22}^{\,\alpha}\,|\,S_{12}^{\,\alpha}\right \}\,=\,\frac{2}{\pi}\frac{|S_{12}^{\,\alpha}|}{(S_{11}^{\,\alpha}\,-\,S_{22}^{\,\alpha})^{2}\,+\,(S_{12}^{\,\alpha})^{2}}\,.
\end{equation}
Mathematically it is more convenient to introduce  $(\xi^{\,\alpha})^{2}\,~=~\,(S_{11}^{\,\alpha}\,-\,S_{22}^{\,\alpha})^{2}\,+\,(S_{12}^{\,\alpha})^{2}$ and determine the probability of finding $S_{11}^{\,\alpha}\,-\,S_{22}^{\,\alpha}$ given  $\xi^{\,\alpha}$.
\begin{equation}\label{frank29}
P \left \{ S_{11}^{\,\alpha}\,-\,S_{22}^{\,\alpha}\,|\,\xi^{\,\alpha} \right \}\,=\,\frac{1}{2\pi}\frac{1}{\sqrt{(\xi^{\,\alpha})^{2}\,-\,(S_{11}^{\,\alpha}\,-\,S_{22}^{\,\alpha})^{2}}}\,.
\end{equation}
The mean values of $S_{11}^{\,\alpha}\,-\,S_{22}^{\,\alpha}$ and $S_{12}^{\,\alpha}$ are zero, hence we predict, rather obviously, hydrostatic pressure. However, notice that we are able to predict the fluctuations away from hydrostatic pressure, and would do more on correlations if one could find a pathway to measuring them. 
\indent Another approach to deal with the system of discrete equations (\ref{frank17}, \ref{frank25}) is to solve it for $S^{\,\alpha}_{ij}$, and then average the solution. This way seems to be feasible at least for the simplest granular systems (e.g. isotropic or periodic arrays) and may provide deep insight into the
origins of the non-gaussian statistics of stress fluctuations \cite{Mueth}. In complex cases this can be accomplished in some approximation, or by using computer simulations.   
By applying Fourier transformation to (\ref{frank17}, \ref{frank25}) one can obtain $S_{ij}({\vec k})$. The macroscopic stress tensor is obtained by averaging over the distribution of angles $\theta^{\,\alpha}$

\begin{equation}\label{frank30}
i\sigma_{11}({\vec k})\,=\,\langle\,S_{11}({\vec k})\,\rangle_{\theta}\,=\,\frac{g_{1}(k_{1}^{3}\,+\,3k_{2}^{2}k_{1})\,+\,g_{2}(k_{2}^{3}\,-\,k_{1}^{2}k_{2})}{|{\vec k}|^{4}}
\end{equation}

\begin{equation}\label{frank31}
i\sigma_{22}({\vec k})\,=\,\langle\,S_{22}({\vec k})\,\rangle_{\theta}\,=\,\frac{g_{2}(k_{2}^{3}\,+\,3k_{1}^{2}k_{2})\,+\,g_{1}(k_{1}k_{2}^{2}\,-\,k_{1}^{3})}{|{\vec k}|^{4}}
\end{equation}

\begin{equation}\label{frank32}
i\sigma_{12}({\vec k})\,=\,\langle\,S_{12}({\vec k})\,\rangle_{\theta}\,=\,\frac{(g_{1}k_{2}-g_{2}k_{1})(k_{2}^{2}\,-\,k_{1}^{2})}{|{\vec k}|^{4}}
\end{equation}
where $|{\vec k}|^{2}\,=\,k_{1}^{2}\,+\,k_{2}^{2}$ and $\sigma_{ij}({\vec r})\,=\,\int\,\sigma_{ij}({\vec k})\,e^{i{\vec k}{\vec r}}\,\mbox{d}^{3}{\vec k}$.  
By doing the inverse Fourier transformation one can see that the macroscopic stress tensor is diagonal.
There must also be constraints on the permitted configurations (due to the absence of tensile forces) which are not so easily expressed, for they affect each grain in the form

\begin{equation}\label{frank33}
S^{\,\alpha}_{ik}\,M^{\,\alpha}_{kl}\,R^{\,\alpha\beta}_{l}\,n^{\,\alpha\beta}_{i}\,>\,0
\end{equation}
which has not yet been put into continuum equations other than $\mbox{Det}\,\sigma\,>\,0$ and $\mbox{Tr}\,\sigma\,>\,0$.

\underline{\sc Discussion} In this Letter we have derived the fundamental equations of stress equilibrium: 

\begin{eqnarray}\label{frank34}
\lefteqn{ \nabla_{j}\,\sigma_{ij}\,+\,\nabla_{j}\nabla_{k}\nabla_{m}\,K_{ijkl}\sigma_{lm}+...\,=\,g_{i}}  \\
& &\mbox{}
 P_{ijk}\sigma_{jk} + \nabla_{j}T_{ijkl}\sigma_{kl} + \nabla_{j}\nabla_{l}U_{ijkl}\sigma_{km} + ... =0\, .\label{frank35}
\end{eqnarray}

In order to solve these equations one needs to know the geometric quantities $K_{ijkl}$, $P_{ijk}$, $T_{ikl}$ and $U_{ijkl}$.
In practice details of the distribution of intergranular contacts are not known in advance, but should be obtained from the deposition history of the system or experimental measurements of two-body correlation functions.

If the system is strongly anisotropic (i.e. there exists a preferred direction characterised by some angle $\phi$) and $\mbox{tan}\,\theta^{\,\alpha}$ has an average value $\mbox{tan}\,\phi$, then eqn.(\ref{frank25}) becomes in the mean-field approximation   
\begin{equation}\label{frank36}
\sigma_{11}\,-\,\sigma_{22}\,=\,2\sigma_{12}\,\mbox{tan} \phi.
\end{equation}
where $\phi$ is the angle of repose. It is known as the Fixed Principal Axes equation \cite{Wittmer}, and has been used with notable effect to solve the problem of the stress distribution in sandpiles.
Explicit mathematical expressions for the 3-D case are more complex, and will be reported elsewhere. The issue of whether the derived system of equations (\ref{frank34}-\ref{frank35}) is robust against the inclusion of   real friciton, softness of grains etc. illuminates the existence of a whole array of fascinating theoretical and experimental problems.
Other important issues which are not addressed in this Letter are that of stress fluctuations and the response of a granular aggregate to external perturbations. In general, cohesionless granular materials are quasistatic or ``fragile''\cite{fragility}, which means that they cannot support certain types of infinitesimal changes in stress without configurational rearrangements.

\indent In conclusion, our theory in its present form gives a simplified, but physical, picture of stress behaviour in cohesionless granular media. Further development is needed to make it a predictive tool which could be able to match experimental findings.

\underline{\sc Acknowledgements}
We acknowledge financial support from Leverhulme Foundation (S. F. E.), Shell (Amsterdam) and Gonville \& Caius College (Cambridge)  (D. V. G.). Invaluable discussions with Prof. Robin Ball are gratefully acknowledged.


\begin{references}
\bibitem{Jaeger}{\it Granular Matter: An
Interdisciplinary Approach}, edited by A. Mehta (Springer-Verlag, New-York, 1993), for review see e.g.  H. M. Jaeger, S. R. Nagel, and R. P. Behringer, Rev. Mod. Phys. {\bf 68}, 1259 (1996).

\bibitem{stress}C. H. Liu et al., Science {\bf 269}, 513 (1995), B. Miller, C. O'Hern and R. P. Behringer, Phys. Rev. Lett. {\bf 77}, 3110 (1996).

\bibitem{Wittmer} J. P. Wittmer, P. Claudin, M. E. Cates,  J. de Physique I (France) {\bf 7} , 39 (1997).

\bibitem{Coppersmith}S. N. Coppersmith et al., Phys. Rev. E, {\bf 53}, 4673 (1996).

\bibitem{Nedderman}R. M. Nedderman, {\it Statics and Kinematics of Granular Materials}, (CUP, Cambridge, 1992).

\bibitem{Wood} D. M. Wood, {\it Soil Behaviour and Critical State Soil Mechanics}, (CUP, Cambridge, 1990).

\bibitem{Grinev} R. C. Ball and D. V. Grinev, {\it cond-mat/9810124}.

\bibitem{Mueth}D. M.~Mueth, H. M.~Jaeger and S. R.~Nagel, Phys. Rev. E, {\bf 57}, 3164 (1998).

\bibitem{fragility} M. E. Cates, J. P. Wittmer, P. Claudin, and J. P. Bouchaud, Phys. Rev. Lett. {\bf 81}, 1841 (1998).
\end{references}
\end{document}